\documentclass[11pt]{article}

\usepackage[T1]{fontenc}
\usepackage[utf8]{inputenc}
\usepackage{lmodern}
\usepackage{microtype}
\usepackage[a4paper,margin=27mm]{geometry}
\usepackage{amsmath,amssymb,mathtools}
\usepackage{booktabs,tabularx,longtable,array,multirow}
\usepackage{graphicx}
\usepackage{xcolor}
\usepackage{enumitem}
\usepackage{listings}
\usepackage{caption}
\usepackage{subcaption}
\usepackage{siunitx}
\usepackage{url}
\usepackage[hidelinks]{hyperref}
\hypersetup{
  pdftitle={FrogBard-512: Design and Experimental Evaluation of a Four-Voice Permutation-Based Hash Function},
  pdfauthor={Victor Duarte Melo},
  pdfsubject={Experimental design and evaluation of a 2048-bit permutation-based hash function},
  pdfkeywords={hash function, permutation, sponge, ARX, tree hashing, AVX2}
}
\usepackage[nameinlink,noabbrev]{cleveref}

\graphicspath{{figures/}}
\setlist{nosep,leftmargin=*}
\definecolor{codebg}{RGB}{246,247,248}
\newcommand{\FB}{FrogBard-512}
\newcommand{\code}[1]{\texttt{#1}}
\newcommand{\ROTL}{\operatorname{ROTL}}
\newcommand{\ROTR}{\operatorname{ROTR}}
\newcommand{\LE}{\operatorname{LE64}}
\newcommand{\xor}{\mathbin{\oplus}}
\newcommand{\boxplusop}{\mathbin{\boxplus}}
\newcommand{\boxminusop}{\mathbin{\boxminus}}

\title{\textbf{FrogBard-512: Design and Experimental Evaluation of a Four-Voice Permutation-Based Hash Function}}
\author{V\'ictor Duarte Melo\\
Independent Researcher, Brazil\\
\texttt{victormeloasm@gmail.com}\\
\url{https://github.com/victormeloasm/FrogBard}}
\date{June 2026}

\begin{document}
\maketitle

\begin{abstract}
This paper presents \FB{}, an experimental 512-bit hash construction built from a custom 2048-bit permutation. The state is arranged as four 512-bit ``voices,'' each containing eight 64-bit words. A round applies public round constants, one of four reproducibly derived byte substitutions, ARX quarter-rounds within each voice, a parity-dependent cross-voice braid, and a fixed lane permutation. Sequential hashing follows a sponge-like design with a 1024-bit rate, a 1024-bit capacity, 128-byte message blocks, and two marker-separated final permutation calls. A distinct tree mode uses 1 MiB leaves, explicit domain separation, root binding, POSIX worker threads, and an AVX2 backend that evaluates four equal-length messages in parallel.

We provide a complete specification, deterministic generation of constants, a portable C11 implementation, a bitsliced table-free profile, inverse reduced-round permutations, fixed vectors, and an empirical evaluation. A 4 h 40 min integrated campaign executed one million randomized API cases, 1,481,187 libFuzzer inputs, one hour of AFL++, multiple sanitizer and Valgrind-family analyses, and two independent 8 GiB output-stream tests. PractRand reported no anomalies in 270 results for either stream; Dieharder reported 224 passed, four weak, and zero failed assessments across both batteries. The full-hash avalanche experiment over one million one-bit perturbations produced a mean of 256.002533 changed output bits and a standard deviation of 11.310838, close to the ideal binomial standard deviation $\sqrt{128}$.

These results support implementation consistency and the absence of obvious statistical defects under the executed tests. They do not establish collision, preimage, second-preimage, indifferentiability, or structural cryptanalytic security. \FB{} remains a research prototype requiring independent analysis.
\end{abstract}


\begin{quote}\small
\textbf{Security status.} \FB{} is a new experimental construction. It has no independent cryptanalysis and is not recommended as a production replacement for SHA-512, SHA-3, BLAKE2, BLAKE3, or other established hashes.
\end{quote}

\section{Introduction}
Cryptographic hash functions provide fixed-length representations of arbitrary messages and are used in integrity checking, signatures, commitment schemes, content addressing, and many protocol constructions. Contemporary designs include iterated compression functions, wide-pipe constructions, and permutation-based sponge functions. The sponge framework is particularly attractive because a single public permutation can support hashing and related modes while separating absorbed data from a hidden capacity portion of the state \cite{bertoni-sponge,keccak-reference,nist-fips202}.

\FB{} is an experimental design intended as a transparent research object rather than as an immediately deployable primitive. Its central structural choice is a 2048-bit permutation organized as four 512-bit voices. The ``voice'' terminology is only a state-layout convention: each voice contains eight 64-bit words, and the four voices are repeatedly mixed by substitution, ARX operations, cross-voice coupling, and lane movement. The sequential hash exposes a 1024-bit rate and retains a 1024-bit capacity. A second domain implements tree hashing for large regular files.

The project has four goals. First, the complete function should be precisely reproducible from a small set of public rules rather than opaque tables. Second, the implementation should support portable scalar, bitsliced, and independent-message SIMD profiles while producing identical digests. Third, reduced-round experimentation should be easy: the implementation exposes exact forward and inverse permutations for every prefix from one to sixteen rounds. Fourth, engineering evidence should be reported honestly and separated from claims of cryptographic security.

\paragraph{Contributions.}
This paper makes the following contributions:
\begin{itemize}
  \item a self-contained specification of the 2048-bit permutation, sequential mode, and domain-separated tree mode;
  \item a deterministic procedure deriving four affine-equivalent AES-based byte substitutions, round constants, and the initial state from public phrases;
  \item a portable C11 implementation, a bitsliced scalar implementation, and an AVX2 four-way backend for equal-length independent messages;
  \item exact inverse permutations for reduced-round testing and fixed conformance vectors;
  \item an integrated evaluation covering compiler diversity, fuzzing, sanitizers, dynamic analysis, concurrency analysis, source coverage, large-stream statistical testing, and reduced-round smoke measurements;
  \item a clear account of the limits of that evidence and a program for independent cryptanalysis.
\end{itemize}

\paragraph{Organization.}
\Cref{sec:design} introduces the design. \Cref{sec:constants,sec:round,sec:modes} specify the constants, permutation, and hashing modes. \Cref{sec:security} discusses intended generic targets and limitations. \Cref{sec:implementation} describes the implementation profiles. \Cref{sec:methodology,sec:results} present the empirical campaign and results. \Cref{sec:discussion,sec:agenda} discuss interpretation and future work.

\section{Design Overview}\label{sec:design}

\subsection{Parameters and state organization}
Let
\[
  W = \mathbb{Z}/2^{64}\mathbb{Z}.
\]
The internal state is
\[
  V=(V_0,V_1,V_2,V_3)\in W^{4\times 8},
  \qquad V_v=(v_{v,0},\ldots,v_{v,7}).
\]
The total state size is $4\cdot8\cdot64=2048$ bits. Voices $V_0$ and $V_1$ form the 1024-bit rate; voices $V_2$ and $V_3$ form the 1024-bit capacity. The digest is 512 bits.

\begin{figure}[t]
  \centering
  \includegraphics[width=\textwidth]{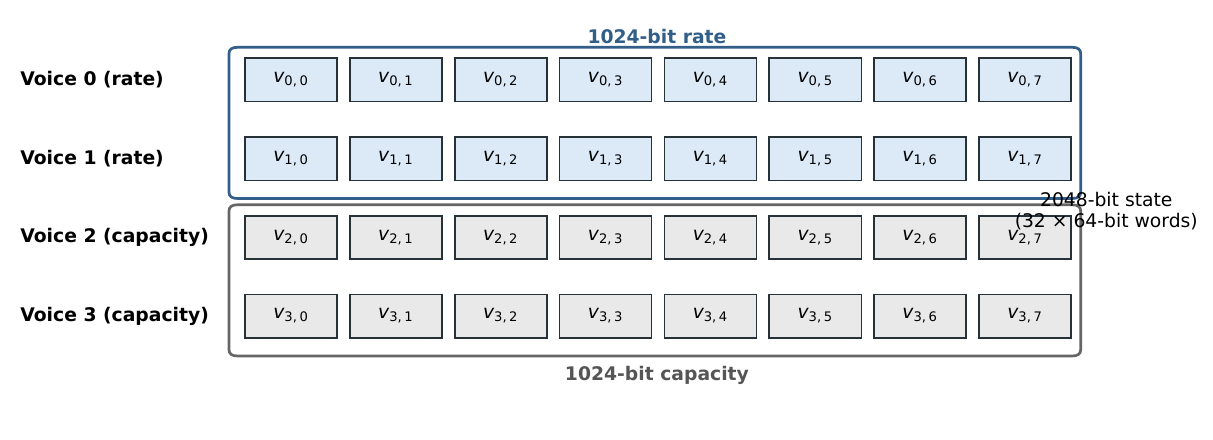}
  \caption{Four-voice organization of the 2048-bit state. Each lane is a 64-bit word.}
  \label{fig:state}
\end{figure}

\begin{table}[t]
\centering
\caption{Core parameters of \FB{} v0.3-experimental.}
\label{tab:parameters}
\begin{tabular}{ll}
\toprule
Parameter & Value \\
\midrule
Digest size & 512 bits (64 bytes) \\
Internal state & 2048 bits (32 64-bit words) \\
Rate & 1024 bits (voices 0 and 1) \\
Capacity & 1024 bits (voices 2 and 3) \\
Sequential block size & 128 bytes \\
Permutation rounds & 16 \\
Additional final calls & 2 full permutation calls \\
External byte order & Little endian \\
Tree leaf payload & At most $2^{20}$ bytes \\
\bottomrule
\end{tabular}
\end{table}

All external words are encoded little endian:
\[
  \LE(b_0,\ldots,b_7)=\sum_{i=0}^{7}b_i2^{8i}.
\]
Addition and subtraction in the state are modulo $2^{64}$, denoted $\boxplusop$ and $\boxminusop$. Bitwise exclusive OR is denoted $\xor$.

\subsection{Round structure}
A round has five layers:
\begin{enumerate}
  \item XOR public round constants into all 32 state words;
  \item apply a voice-dependent byte substitution to every state byte;
  \item apply four ARX quarter-rounds independently within each voice;
  \item apply a parity-dependent cross-voice braid;
  \item apply parity-dependent lane permutations.
\end{enumerate}

\begin{figure}[t]
  \centering
  \includegraphics[width=\textwidth]{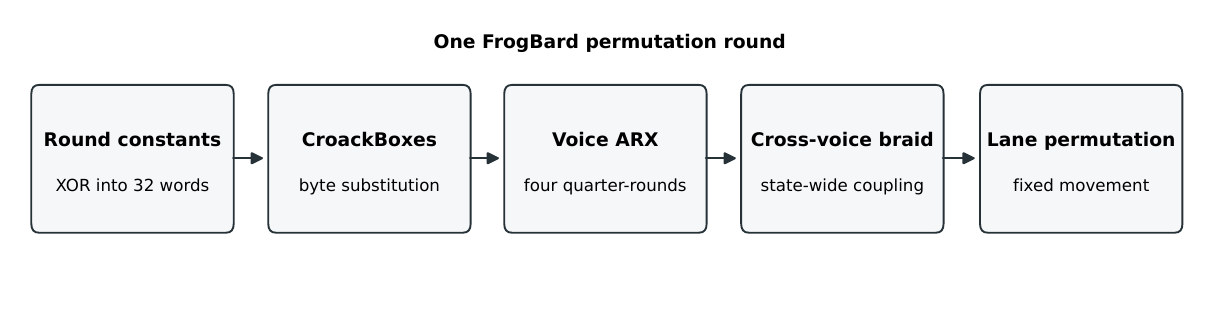}
  \caption{High-level sequence of one permutation round. The first three layers are evaluated per voice; the braid couples the complete state.}
  \label{fig:round-pipeline}
\end{figure}

The full permutation is the composition of rounds $0$ through $15$. Each component is bijective, which enables an exact inverse for any reduced-round prefix.

\section{Reproducible Constants and CroackBoxes}\label{sec:constants}

\subsection{Public phrases}
The construction uses four exact ASCII byte strings without a trailing newline:
\begin{enumerate}
  \item \code{Libertas per Croack!}
  \item \code{Presunto}
  \item \code{Floquinha}
  \item \code{In Frog we Trust!}
\end{enumerate}
They are transparent public seeds, not keys, passwords, or secret material.

\subsection{Project-defined 64-bit mixer}
For domain bytes $D$, phrase bytes $P$, and counter $c\in W$, define \(\operatorname{Mix64}(D,P,c)\) by the following recurrence, with multiplication modulo $2^{64}$:
\begin{align*}
  h &\leftarrow 1469598103934665603,\\
  h &\leftarrow (h\xor b)\cdot1099511628211 \quad \text{for each } b\in D,\\
  h &\leftarrow (h\xor 0)\cdot1099511628211,\\
  h &\leftarrow (h\xor b)\cdot1099511628211 \quad \text{for each } b\in P,\\
  h &\leftarrow (h\xor ((c\gg 8i)\mathbin{\&}255))\cdot1099511628211
       \quad \text{for } i=0,\ldots,7.
\end{align*}
The explicit zero byte separates the domain and phrase. This function is a deterministic generator component, not a cryptographic hash claim.

The mixer output seeds SplitMix64 expansion:
\begin{align*}
 s' &= s\boxplusop\mathtt{9e3779b97f4a7c15},\\
 z_1 &= (s'\xor(s'\gg30))\cdot\mathtt{bf58476d1ce4e5b9},\\
 z_2 &= (z_1\xor(z_1\gg27))\cdot\mathtt{94d049bb133111eb},\\
 z_3 &= z_2\xor(z_2\gg31).
\end{align*}

\subsection{CroackBox definition}
Let $S_{\mathrm{AES}}$ be the AES S-box \cite{daemen-rijmen}. For phrase index $j$, the generator derives input and output bit permutations $p^{\mathrm{in}}_j,p^{\mathrm{out}}_j$ and bytes $a_j,b_j$. The resulting CroackBox is
\[
  S_j(x)=\operatorname{BP}_{p^{\mathrm{out}}_j}
  \left(S_{\mathrm{AES}}\left(\operatorname{BP}_{p^{\mathrm{in}}_j}(x\xor a_j)\right)\right)\xor b_j,
\]
where \(\operatorname{BP}_{p}\) permutes the eight bits of a byte. Candidates are rejected if any $x$ satisfies either $S_j(x)=x$ or $S_j(x)=x\xor\mathtt{ff}$.

Because the input and output maps are affine bijections, each CroackBox is affine-equivalent to the AES S-box. Bijectivity, differential uniformity 4, nonlinearity, and algebraic degree are therefore preserved under the transformation \cite{nyberg}. These isolated S-box properties do not establish the security of the complete permutation.

\begin{table}[t]
\centering
\caption{Exact CroackBox metadata. Permutations list the input bit selected for each output bit.}
\label{tab:croackboxes}
\small
\begin{tabularx}{\textwidth}{cXllcc}
\toprule
$j$ & Phrase & $p^{\mathrm{in}}_j$ & $p^{\mathrm{out}}_j$ & $a_j$ & $b_j$ \\
\midrule
0 & Libertas per Croack! & 3,6,5,7,1,0,4,2 & 7,6,0,3,4,5,2,1 & A7 & 71 \\
1 & Presunto & 3,7,6,0,2,4,1,5 & 3,0,2,5,6,4,1,7 & 72 & 43 \\
2 & Floquinha & 5,3,2,4,7,6,1,0 & 6,7,3,4,1,5,0,2 & 8F & 9E \\
3 & In Frog we Trust! & 5,7,4,2,3,1,6,0 & 1,3,6,2,7,0,4,5 & 8A & 76 \\
\bottomrule
\end{tabularx}
\end{table}
The accepted candidate counters are 11, 6, 10, and 18, respectively.

\subsection{Round constants and initial state}
Let
\[
  \sigma_0=\mathtt{46524f4742415244},
\]
whose hexadecimal representation spells \code{FROGBARD}. The four phrases and a versioned domain string seed a SplitMix64 stream. For round $r\in\{0,\ldots,15\}$, voice $v\in\{0,\ldots,3\}$, and lane $\ell\in\{0,\ldots,7\}$, one stream word $x$ is transformed into
\[
 \mathrm{RC}_{r,v,\ell}=x
 \xor \left(\mathtt{9e3779b97f4a7c15}(r+1)\right)
 \xor \ROTL_{\ell+1}\left(\mathtt{d1b54a32d192ed03}(v+1)\right).
\]
The generated table contains 512 round-constant words. The initial state is generated separately from a versioned IV domain and the corresponding phrase for each voice. The generator script is normative for constant reproduction; the generated header has SHA-256 digest
\begin{center}
\code{7a4d1b7bdd9aad454ce7bb513d9126608695afea2db33b5ab26a900608dabd9d}.
\end{center}

\section{The 2048-bit Permutation}\label{sec:round}

\subsection{ARX quarter-round}
For $(a,b,c,d)\in W^4$ and rotations $(r_0,r_1,r_2,r_3)$, define
\begin{align*}
 a&\leftarrow a\boxplusop b, & d&\leftarrow\ROTL_{r_0}(d\xor a),\\
 c&\leftarrow c\boxplusop d, & b&\leftarrow\ROTL_{r_1}(b\xor c),\\
 a&\leftarrow a\boxplusop b, & d&\leftarrow\ROTL_{r_2}(d\xor a),\\
 c&\leftarrow c\boxplusop d, & b&\leftarrow\ROTL_{r_3}(b\xor c).
\end{align*}
This is a 256-bit bijection composed of modular addition, XOR, and fixed rotation, in the general ARX tradition exemplified by Salsa20 and ChaCha \cite{bernstein-chacha}.

The exact inverse applies the primitive inverses in reverse order:
\begin{align*}
 b&\leftarrow\ROTR_{r_3}(b)\xor c, & c&\leftarrow c\boxminusop d,\\
 d&\leftarrow\ROTR_{r_2}(d)\xor a, & a&\leftarrow a\boxminusop b,\\
 b&\leftarrow\ROTR_{r_1}(b)\xor c, & c&\leftarrow c\boxminusop d,\\
 d&\leftarrow\ROTR_{r_0}(d)\xor a, & a&\leftarrow a\boxminusop b.
\end{align*}

\subsection{Within-voice schedule}
The four base rotation tuples are
\begin{align*}
 R_0&=(17,29,41,53), & R_1&=(23,31,47,59),\\
 R_2&=(11,37,43,57), & R_3&=(19,27,45,61).
\end{align*}
Voice $v$ in round $r$ selects $R_{(v+r)\bmod4}=(\rho_0,\rho_1,\rho_2,\rho_3)$. For voice $X=(x_0,\ldots,x_7)$, the following quarter-rounds are executed in order:
\begin{enumerate}
  \item $Q_{\rho_0,\rho_1,\rho_2,\rho_3}(x_0,x_1,x_2,x_3)$;
  \item $Q_{\rho_1,\rho_2,\rho_3,\rho_0}(x_4,x_5,x_6,x_7)$;
  \item $Q_{\rho_2,\rho_3,\rho_0,\rho_1}(x_0,x_5,x_2,x_7)$;
  \item $Q_{\rho_3,\rho_0,\rho_1,\rho_2}(x_4,x_1,x_6,x_3)$.
\end{enumerate}
The last two calls cross-link the initial lane quartets.

\subsection{Cross-voice braid}
Define
\begin{align*}
 A&=(7,13,19,29,37,43,53,61),\\
 B&=(11,17,23,31,41,47,55,59).
\end{align*}
Indices are reduced modulo 8, and assignments are sequential. For even rounds and $i=0,\ldots,7$:
\begin{align*}
 v_{0,i}&\leftarrow v_{0,i}\boxplusop\ROTL_{A_i}(v_{1,i+1}),\\
 v_{2,i}&\leftarrow v_{2,i}\xor\ROTL_{B_i}(v_{0,i+3}),\\
 v_{3,i}&\leftarrow v_{3,i}\boxplusop\ROTL_{A_{i+2}}(v_{2,i+5}),\\
 v_{1,i}&\leftarrow v_{1,i}\xor\ROTL_{B_{i+4}}(v_{3,i+7}).
\end{align*}
For odd rounds:
\begin{align*}
 v_{3,i}&\leftarrow v_{3,i}\boxplusop\ROTL_{B_i}(v_{2,i+2}),\\
 v_{1,i}&\leftarrow v_{1,i}\xor\ROTL_{A_i}(v_{3,i+4}),\\
 v_{0,i}&\leftarrow v_{0,i}\boxplusop\ROTL_{B_{i+3}}(v_{1,i+6}),\\
 v_{2,i}&\leftarrow v_{2,i}\xor\ROTL_{A_{i+5}}(v_{0,i+1}).
\end{align*}
The braid is inverted by processing $i$ from 7 down to 0 and undoing each assignment in reverse order.

\subsection{Lane permutations and full round}
Each voice applies a fixed parity-dependent lane permutation. The exact maps are listed in \Cref{app:permutations}. Let $K_r$ denote round-constant XOR, $S_r$ the four voice-dependent CroackBoxes, $M_r$ the four within-voice ARX mixers, $B_r$ the braid, and $\Pi_r$ the lane movement. Then
\[
 R_r=\Pi_r\circ B_r\circ M_r\circ S_r\circ K_r,
 \qquad
 P=R_{15}\circ\cdots\circ R_0.
\]
Every component is bijective. Therefore every round and the full 16-round permutation are bijections on $\{0,1\}^{2048}$.

\section{Hashing Modes}\label{sec:modes}

\subsection{Sequential mode}
Initialization copies the generated IV and applies one full permutation:
\[
 V^{(0)}=P(\mathrm{IV}).
\]
A 128-byte block is decoded into words $m_0,\ldots,m_{15}$ and injected into the rate:
\[
 v_{0,i}\leftarrow v_{0,i}\xor m_i,
 \qquad
 v_{1,i}\leftarrow v_{1,i}\xor m_{i+8},
 \qquad i=0,\ldots,7,
\]
followed by $V\leftarrow P(V)$. Capacity voices are not directly XORed with message words.

For $\ell\in\{0,\ldots,127\}$ buffered bytes, the final block is zero-filled and encoded as
\[
 B[\ell]\leftarrow B[\ell]\xor\mathtt{0b},
 \qquad
 B[127]\leftarrow B[127]\xor\mathtt{80}.
\]
A final block is always absorbed, including when the message length is a multiple of 128 bytes. After the padded block and its permutation, the 128-bit byte length $(L_{\mathrm{hi}},L_{\mathrm{lo}})$ is injected into $v_{2,6},v_{2,7}$. Two public final-domain words are injected into $v_{3,6},v_{3,7}$:
\[
 \mathtt{496e2046726f6720},\qquad \mathtt{7765205472757374}.
\]
Two marker-separated final calls follow:
\[
 v_{3,0}\leftarrow v_{3,0}\xor\mathtt{21};\ V\leftarrow P(V),
\]
\[
 v_{3,1}\leftarrow v_{3,1}\xor\mathtt{2100};\ V\leftarrow P(V).
\]
The output is voice 0 serialized little endian.

\subsection{Tree mode}
Tree mode is a distinct domain and normally produces a digest different from sequential mode, even for a one-leaf file. The leaf payload is $C=2^{20}$ bytes. For chunk $C_i$, index $i$, and length $|C_i|$:
\[
 L_i=H_{\mathrm{FB}}(D_L\parallel\LE(i)\parallel\LE(|C_i|)\parallel C_i),
\]
where $D_L$ is the exact 16-byte string \code{FrogBardTreeL1\textbackslash0\textbackslash0}. The empty file has one zero-length leaf.

A parent message is exactly 160 bytes:
\[
 D_P\parallel\LE(\ell)\parallel\LE(k)\parallel L\parallel R,
\]
where $D_P$ is \code{FrogBardTreeP1\textbackslash0\textbackslash0}, $\ell$ is the zero-based reduction level, $k\in\{1,2\}$ is the child count, and $R$ is all zero when $k=1$.

After reduction to one internal root $R_{\mathrm{int}}$, the bound root message is
\[
 D_R\parallel\LE(F)\parallel\LE(C)\parallel\LE(N)\parallel\LE(h)\parallel R_{\mathrm{int}},
\]
where $F$ is the file size, $N$ the number of leaves, $h$ the number of reduction levels, and $D_R$ is \code{FrogBardTreeR1\textbackslash0\textbackslash0}.

\begin{figure}[t]
  \centering
  \includegraphics[width=\textwidth]{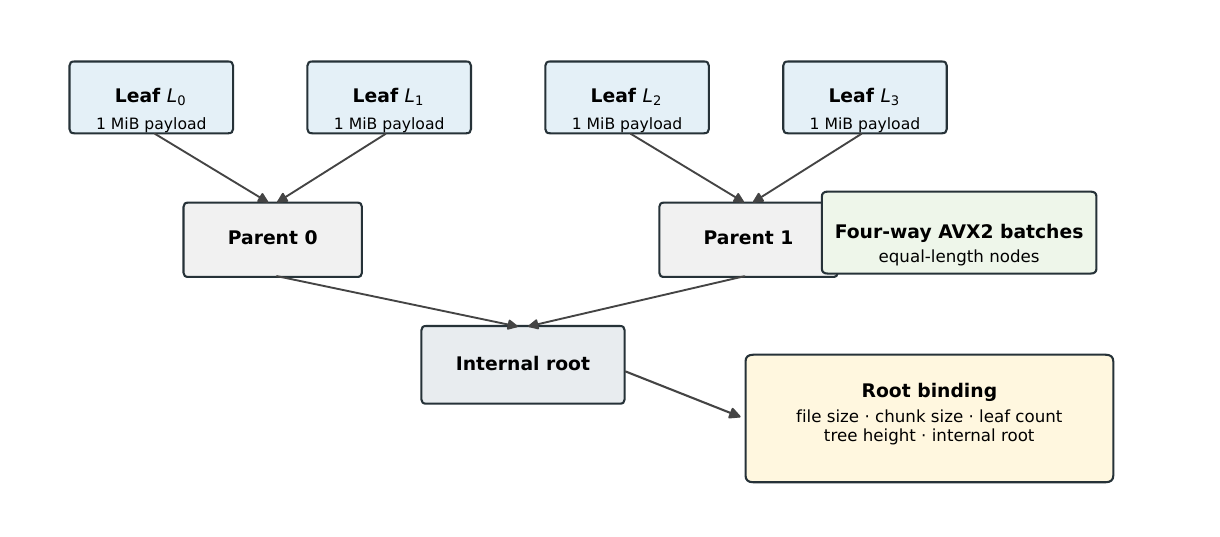}
  \caption{Domain-separated tree mode. The final root binds file size, chunk size, leaf count, tree height, and the internal root digest.}
  \label{fig:tree}
\end{figure}

Workers use \code{pread()} and claim groups of up to four leaves. Full groups with equal message lengths are processed by the AVX2 four-way backend; incomplete groups use scalar calls.

\section{Security Goals and Limitations}\label{sec:security}

For a sponge-like construction with capacity $c=1024$ and output length $n=512$, an ideal-permutation heuristic suggests generic targets of approximately
\begin{align*}
 \text{collision work} &\approx 2^{\min(n/2,c/2)}=2^{256},\\
 \text{preimage work} &\approx 2^{\min(n,c)}=2^{512},\\
 \text{second-preimage work} &\approx 2^{\min(n,c)}=2^{512}.
\end{align*}
These are design targets in an idealized model, not proven security levels for the concrete permutation.

The construction has several explicit caveats:
\begin{itemize}
  \item the four CroackBoxes are affine-equivalent variants of one S-box rather than independent random permutations;
  \item no wide-trail lower bound is supplied for active S-boxes across the ARX, braid, and lane layers;
  \item no formal upper bounds are supplied for differential or linear probabilities of the complete round function;
  \item the constant generator is deterministic and transparent but intentionally not claimed to be cryptographically pseudorandom;
  \item the choice of sixteen rounds has not yet been justified by the strongest attacks on reduced variants;
  \item the default table implementation performs data-dependent S-box accesses;
  \item the tree construction requires independent analysis of its domain separation, prefix encoding, unary-parent rule, and root binding.
\end{itemize}

A large state does not compensate for a structural weakness. Differential trails, rotational relations, invariant subspaces, rebound-style attacks, algebraic shortcuts, meet-in-the-middle decompositions, or unexpected symmetries could reduce the effective security well below the generic bounds.

\section{Implementation Profiles}\label{sec:implementation}

\subsection{Portable scalar implementation}
The sequential core is written in C11 and uses no dynamic allocation. The state and input buffer are aligned to 64 bytes. The default forward path uses four 256-byte substitution tables, for a total S-box footprint of 1 KiB. The principal hot data are the 256-byte state, the 1 KiB S-box set, and the 4 KiB round-constant table.

\subsection{Bitsliced scalar implementation}
A table-free profile applies an orthogonal bitslice transform, a Boolean AES S-box circuit based on Boyar--Peralta logic minimization \cite{boyar-peralta}, and the generated affine maps. It produces exactly the same digests as the table profile. This removes message-dependent scalar S-box addresses, although it does not turn the hash into a keyed primitive or establish side-channel security for arbitrary compositions.

\subsection{AVX2 four-way implementation}
The SIMD backend hashes four equal-length independent messages. Each 256-bit vector contains the corresponding 64-bit word of four separate states:
\[
 X_{v,\ell}=(v^{(0)}_{v,\ell},v^{(1)}_{v,\ell},v^{(2)}_{v,\ell},v^{(3)}_{v,\ell}).
\]
This is independent-message parallelism: SIMD lanes do not represent pieces of a single hash state. Native tree mode uses the backend for leaf and parent batches.

\subsection{Conformance surface}
The implementation provides one-shot, incremental, four-way, regular-file tree, reduced-round, and inverse-reduced-round APIs. Conformance requires exact byte order, constants, operation ordering, padding markers, length injection, tree domains, and output extraction. Selected fixed vectors appear in \Cref{app:vectors}.

\section{Experimental Methodology}\label{sec:methodology}

\subsection{Campaign scope}
The integrated \code{insane} profile ran on 19 June 2026. It was designed to examine five evidence classes:
\begin{enumerate}
  \item build and backend conformance across compilers and optimization profiles;
  \item memory, undefined-behavior, leak, and concurrency diagnostics;
  \item randomized and coverage-guided dynamic testing;
  \item reduced-round smoke analysis;
  \item large-stream general-purpose statistical testing.
\end{enumerate}

The campaign ran for 16,842 seconds (4 h 40 min 42 s) and created approximately 17 GiB of artifacts. The host used Ubuntu 26.04 LTS, Linux 7.0.0-22, an AMD Ryzen 9 5950X with 16 cores and 32 threads, and approximately 121 GiB of reported physical memory. The primary toolchain was Clang/LLD 21.1.8; GCC 15.2.0 was also available. Dynamic tools included Valgrind 3.26.0, AFL++ 4.33c, PractRand 0.96, Dieharder 3.31.1, and LLVM coverage tools.

\subsection{Result accounting}
The raw harness summary contained 250 \code{PASS}, nine \code{FAIL}, and six \code{WARN} records. Exit codes and result files show a more informative classification:
\begin{itemize}
  \item 250 records completed with exit status zero;
  \item two PractRand runs completed the requested 8 GiB test and reported no anomalies, but the pipeline returned 141 because an infinite producer received SIGPIPE after PractRand closed the pipe;
  \item one Cppcheck record requires review of a possible unary-parent pointer dereference and was contaminated by third-party PractRand findings because the scan scope was too broad;
  \item six benchmark or resource-limit tests ended with exit 130 after manual interruption and are incomplete rather than observed cryptographic failures.
\end{itemize}

\begin{figure}[t]
  \centering
  \includegraphics[width=0.82\textwidth]{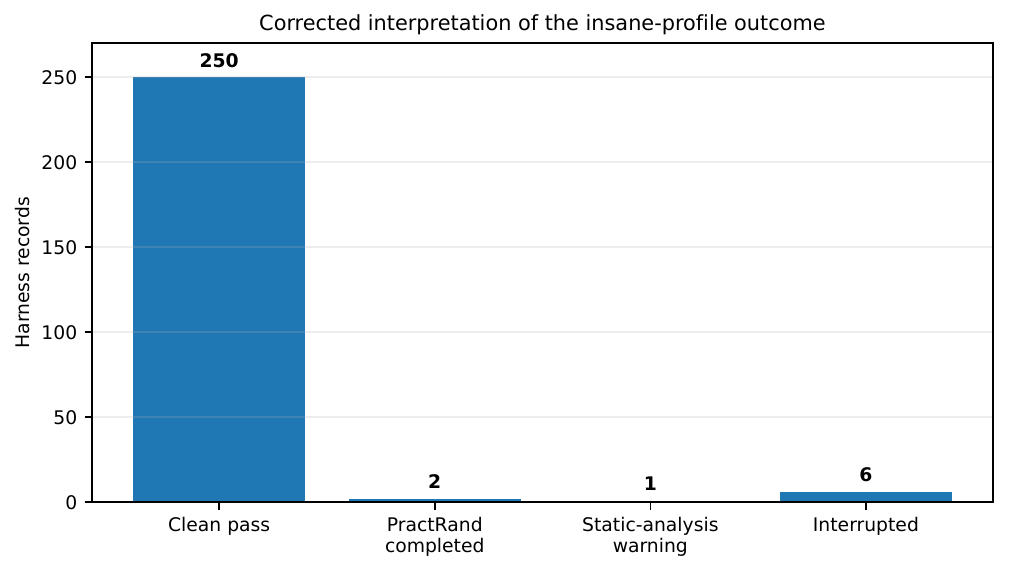}
  \caption{Corrected interpretation of the raw harness result.}
  \label{fig:outcome}
\end{figure}

\subsection{Generated streams}
The counter stream hashes 64-byte messages containing a 64-bit counter and a second counter-derived word. The differential stream emits
\[
 H(M_i)\xor H(M_i\xor e_{i\bmod512}),
\]
where the flipped input bit cycles through the 512-bit input. Each stored stream contains $2^{33}$ bytes (8 GiB).

\subsection{Reduced-round experiments}
State avalanche experiments select random 2048-bit states, flip one input bit, evaluate a reduced-round prefix, and count output differences. Separate scans sample linear masks and one-bit differentials. Exact inverse tests check
\[
 P_r^{-1}(P_r(x))=x
\]
for 16,384 sampled states per round count, $r=1,\ldots,16$. Deliberately truncated 32-bit collision tests provide only a harness sanity check; such collisions are expected near the birthday scale and do not address the full 512-bit digest.

\section{Experimental Results}\label{sec:results}

\subsection{Build, conformance, and reproducibility}
Scalar builds succeeded under Clang at \code{-O0}, \code{-O1}, \code{-O2}, \code{-O3}, \code{-Os}, and \code{-Oz}. Native, portable, constant-time, and hardened profiles produced the expected self-test outputs. Fixed vectors, one-shot versus incremental hashing, arbitrary update segmentation, padding boundaries, scalar versus four-way hashing, and tree determinism across thread counts passed in the completed tests. Regenerated tables were byte-identical to the committed header, and a repeated native build produced a bit-for-bit identical binary under the recorded environment.

\subsection{Memory safety and concurrency}
Completed AddressSanitizer, LeakSanitizer, UndefinedBehaviorSanitizer, and ThreadSanitizer runs did not report an implementation defect. Valgrind Memcheck reported zero errors and no leaks for the tested self-test and streaming paths. Helgrind and DRD reported zero detected synchronization errors in the exercised tree mode. These tools are complementary but incomplete: absence of a diagnostic is not a proof of memory or concurrency safety.

\subsection{Randomized API testing and fuzzing}
A dedicated stress harness executed 1,000,000 randomized API cases, comparing one-shot and fragmented streaming calls and exercising the four-way backend. No mismatch or crash was reported. The same run measured full-hash avalanche over one million one-bit perturbations:
\begin{table}[h]
\centering
\caption{Full-hash avalanche summary over one million cases.}
\label{tab:full-avalanche}
\begin{tabular}{lr}
\toprule
Metric & Observed value \\
\midrule
Mean changed output bits & 256.002533 \\
Minimum & 201 \\
Maximum & 314 \\
Standard deviation & 11.310838 \\
Ideal binomial standard deviation & $\sqrt{512\cdot\tfrac12\cdot\tfrac12}=11.313708$ \\
\bottomrule
\end{tabular}
\end{table}

The mean and standard deviation are close to those of independent fair output bits. This is useful evidence against gross diffusion defects but is not a bound on differential probability.

libFuzzer ran for 3,601 seconds, executed 1,481,187 inputs, reached 527 coverage units and 1,541 feature units in its final log, and reported no crash. Peak resident memory was 432 MiB. AFL++ ran for one hour, found 21 new corpus items, reported 63.27\% instrumentation coverage, and saved zero crashes and zero timeouts.

\subsection{Reduced-round diffusion and sampled biases}
\begin{figure}[t]
  \centering
  \includegraphics[width=0.88\textwidth]{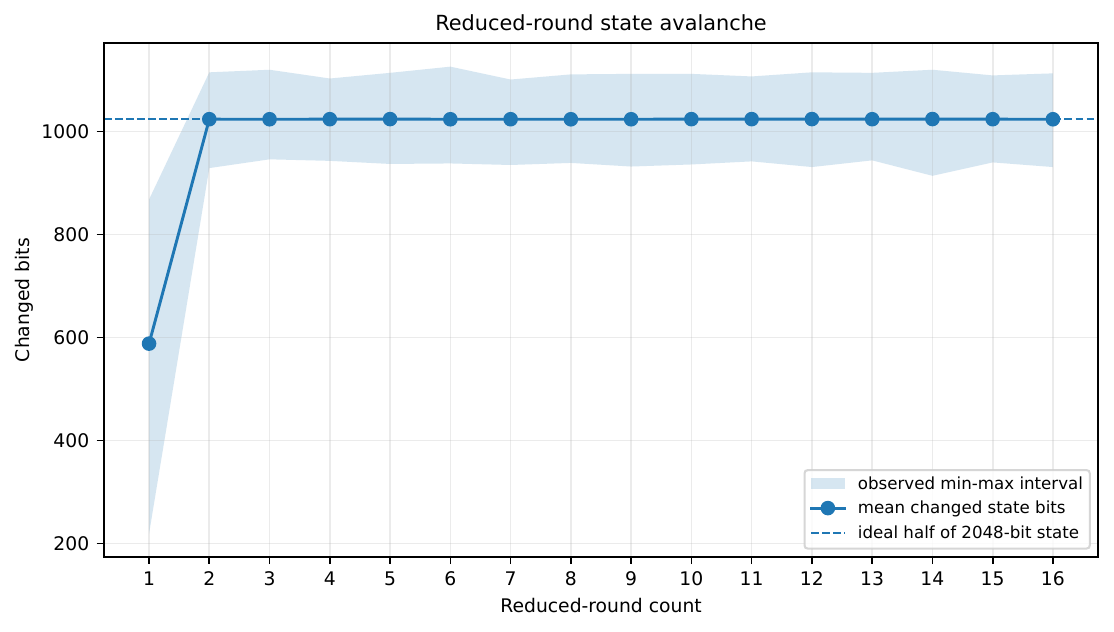}
  \caption{Reduced-round state avalanche. The shaded region is the sampled minimum-to-maximum interval, not a confidence interval.}
  \label{fig:round-avalanche}
\end{figure}

\Cref{fig:round-avalanche} shows the 2048-bit state avalanche results. One round changes approximately 588 bits on average. By two rounds the mean reaches approximately 1024 bits and remains near half the state through sixteen rounds.

\begin{table}[t]
\centering
\caption{Selected reduced-round smoke measurements. Bias is absolute deviation from one half.}
\label{tab:reduced-scans}
\small
\resizebox{\textwidth}{!}{%
\begin{tabular}{rrrrrr}
\toprule
Rounds & Avalanche mean & Rotational distance & Linear bias & Differential bit bias & 32-bit collision attempts \\
\midrule
1  & 588.003  & 1024.083 & 0.00631714 & 0.50000000 & 72,464 \\
2  & 1023.946 & 1024.199 & 0.00677490 & 0.00994873 & 141,542 \\
4  & 1023.969 & 1023.623 & 0.00694275 & 0.00971985 & 77,927 \\
8  & 1023.821 & 1024.229 & 0.00675964 & 0.01010132 & 108,565 \\
12 & 1024.169 & 1023.916 & 0.00663757 & 0.01016235 & 109,640 \\
16 & 1023.845 & 1024.075 & 0.00683594 & 0.01098633 & 69,279 \\
\bottomrule
\end{tabular}%
}
\end{table}

The one-round differential result contains a completely biased output bit for one sampled input difference, which is expected before full diffusion and directly shows that one round is distinguishable. From two rounds onward, the strongest sampled bit biases in this campaign were near one percent. These are finite-sample scans over selected differences, output bits, and masks; they neither locate optimal trails nor bound unsampled behavior.

All inverse tests passed for 16,384 samples per round count. Four simple equality-oriented invariant patterns did not retain equal voices or equal lanes after the tested reduced-round prefixes. This does not rule out more complex affine, nonlinear, or structured invariant subspaces.

\subsection{Large-stream statistical testing}
PractRand 0.96 evaluated each 8 GiB stream with its core battery and standard 64-bit folding. The counter stream completed in 837 seconds and the differential stream in 1,611 seconds. Each result contained the statement ``no anomalies in 270 test result(s).''

Dieharder reported the outcomes shown in \Cref{tab:dieharder}. Occasional \code{WEAK} p-values are expected when many tests are performed and should be interpreted through independent repetitions rather than as isolated evidence.

\begin{table}[h]
\centering
\caption{Dieharder battery classifications.}
\label{tab:dieharder}
\begin{tabular}{lrrr}
\toprule
Stream & Passed & Weak & Failed \\
\midrule
Counter-derived output & 111 & 3 & 0 \\
One-bit differential output & 113 & 1 & 0 \\
\textbf{Total} & \textbf{224} & \textbf{4} & \textbf{0} \\
\bottomrule
\end{tabular}
\end{table}

Byte-level measurements appear in \Cref{tab:streams}. Both stored files were incompressible under ENT's estimate, had arithmetic means near 127.5, and had serial correlations close to zero.

\begin{table}[h]
\centering
\caption{Selected measurements for the two stored 8 GiB streams.}
\label{tab:streams}
\begin{tabular}{lrr}
\toprule
Metric & Counter stream & Differential stream \\
\midrule
Bytes & 8,589,934,592 & 8,589,934,592 \\
Entropy (bits/byte) & 7.999999981 & 7.999999979 \\
Chi-square, 256 bins & 229.598844 & 251.431989 \\
ENT exceedance probability & 87.17\% & 55.14\% \\
Arithmetic byte mean & 127.5013 & 127.5008 \\
Serial correlation & -0.000018002 & -0.000004567 \\
Estimated optimum compression & 0\% & 0\% \\
\bottomrule
\end{tabular}
\end{table}

General-purpose randomness batteries can expose obvious regularities, but a cryptographic hash can pass them and still be catastrophically weak. These results should therefore be read as smoke tests only \cite{nist-sp80022,testu01}.

\subsection{Coverage and runtime profile}
LLVM coverage reached 80.50\% of lines, 66.10\% of branches, and 96.30\% of functions overall. The core had substantially higher branch and line coverage than CLI and tree error paths.

\begin{figure}[t]
  \centering
  \includegraphics[width=0.86\textwidth]{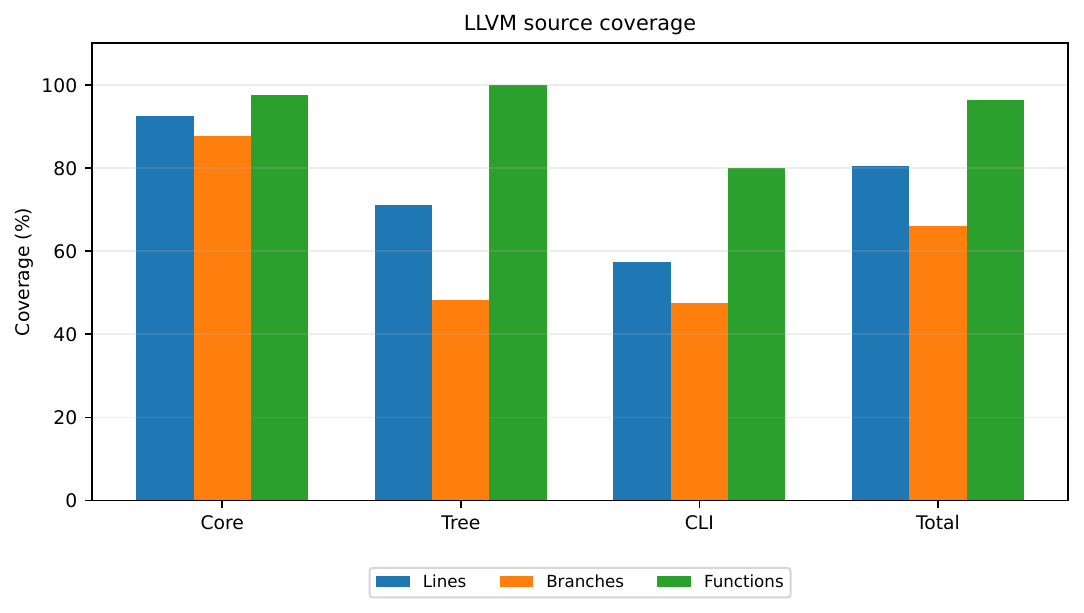}
  \caption{LLVM source coverage from the integrated run.}
  \label{fig:coverage}
\end{figure}

\begin{figure}[t]
  \centering
  \includegraphics[width=0.92\textwidth]{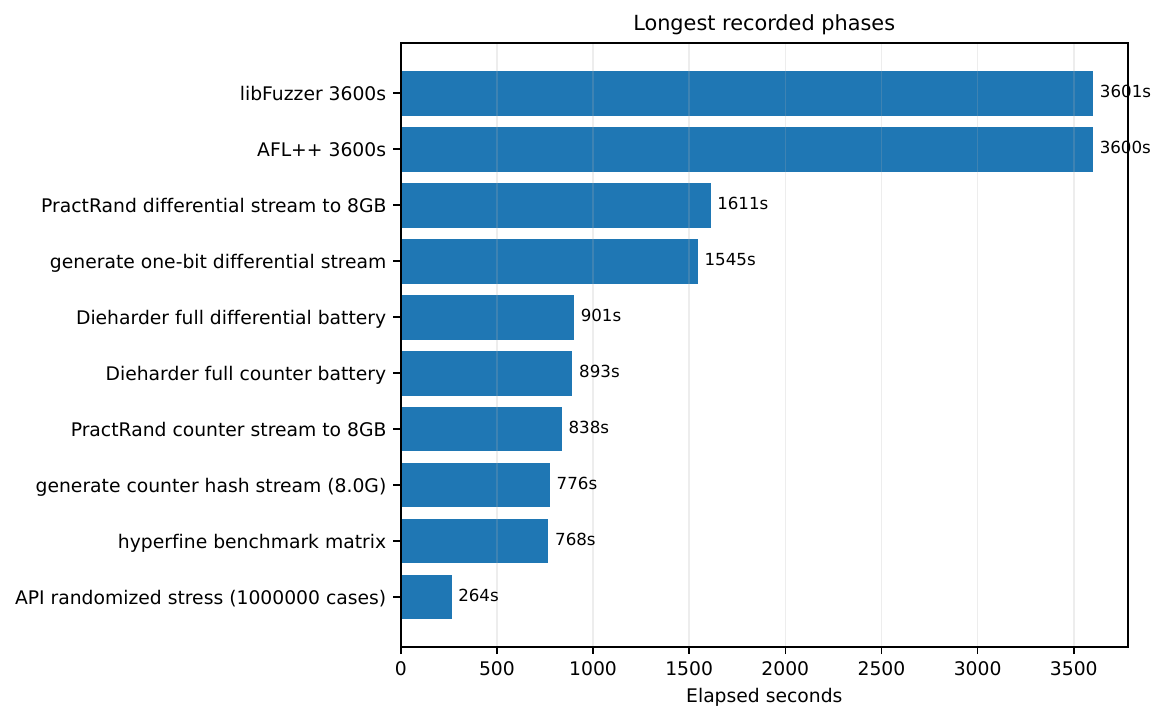}
  \caption{Longest recorded phases. The two PractRand pipeline records are raw harness failures despite successful statistical completion; the Hyperfine stage was interrupted.}
  \label{fig:runtime}
\end{figure}

The interrupted benchmark stage prevents a clean new throughput claim from this campaign. An earlier local snapshot measured approximately 1.17 s for sequential table hashing, 2.15 s for scalar bitslicing, and 0.11 s for 8-thread AVX2 tree hashing on a 64 MiB zero-filled file. Those values are environment-specific and compare different digest domains; they are not portable performance guarantees.

\section{Discussion}\label{sec:discussion}

\subsection{What the evidence supports}
The completed tests support the following restricted claims:
\begin{itemize}
  \item the tested implementations are internally consistent across scalar, bitsliced, and AVX2 paths;
  \item fixed vectors, streaming segmentation, padding boundaries, and tree determinism behave as specified in the exercised cases;
  \item multiple dynamic analyzers found no memory, leak, undefined-behavior, or concurrency defect in the tested paths;
  \item fuzzers and randomized stress tests found no crash or digest mismatch during the recorded runs;
  \item the sampled avalanche and large output streams do not exhibit gross non-random behavior under the executed measurements.
\end{itemize}

\subsection{What the evidence does not support}
The campaign does not establish:
\begin{itemize}
  \item 256-bit collision resistance or 512-bit preimage/second-preimage resistance;
  \item indifferentiability from a random oracle;
  \item resistance to differential, linear, rotational, rebound, boomerang, integral, impossible-differential, algebraic, meet-in-the-middle, or invariant-subspace attacks;
  \item a sufficient 16-round security margin;
  \item production readiness or suitability for standardization;
  \item side-channel safety in uses outside the stated unkeyed hashing context.
\end{itemize}

\subsection{Harness and implementation findings}
The evaluation also identified useful engineering improvements:
\begin{enumerate}
  \item accept SIGPIPE from an unbounded stream producer when a bounded PractRand consumer exits successfully;
  \item document disk requirements accurately: the large profile stores two 8 GiB streams;
  \item classify exit 130 as interrupted rather than failed;
  \item exclude third-party PractRand source from core Cppcheck accounting;
  \item isolate performance stages so they can be resumed independently;
  \item rewrite the unary-parent pointer condition in an explicitly analyzer-friendly form and add a direct unit test for that path.
\end{enumerate}
The Cppcheck report is not treated here as a confirmed memory defect. The relevant code path appears to condition the right-child copy on the child count, but a clearer formulation is appropriate because static analyzers could not prove the invariant.

\section{Recommended Cryptanalytic Program}\label{sec:agenda}
The next research priority is structured cryptanalysis rather than larger volumes of generic randomness tests:
\begin{enumerate}
  \item encode reduced rounds as bit-vector constraints and search differential and linear trails with SAT, SMT, MILP, or constraint programming;
  \item derive active-CroackBox lower bounds across the ARX, braid, and lane layers;
  \item model modular-addition carries and rotational-XOR/addition relations;
  \item search for invariant subspaces, parity-induced symmetries, fixed points, short cycles, and slide relations;
  \item measure algebraic degree growth and equation sparsity;
  \item investigate rebound, boomerang, rectangle, integral, and impossible-differential distinguishers;
  \item study meet-in-the-middle and splice-and-cut decompositions around alternating round parity;
  \item analyze the tree domain for prefix-freeness, multicollisions, herding, and unary-parent edge cases;
  \item publish reduced-round challenges and report the strongest attacks before freezing a non-experimental version.
\end{enumerate}

\section{Reproducibility and Artifact Availability}
The reference implementation, generator, test vectors, reduced-round analysis utility, and harness are available at
\begin{center}
\url{https://github.com/victormeloasm/FrogBard}.
\end{center}
The source package accompanying this paper contains the LaTeX manuscript, all figures used by the paper, machine-readable reduced-round data, and a figure-generation script. The arXiv-specific archive omits compiled PDFs, logs, auxiliary files, and unrelated repository artifacts.

\section{Conclusion}
\FB{} is a fully specified experimental hash construction with a 2048-bit four-voice permutation, a 1024-bit rate and capacity, four reproducibly derived AES-affine CroackBoxes, ARX voice mixing, a cross-voice braid, and a separately domain-separated tree mode. The implementation supports portable scalar, bitsliced scalar, and AVX2 four-way execution, together with exact reduced-round inversion and fixed vectors.

The integrated evaluation found no completed implementation or statistical failure: one million randomized API cases, two one-hour fuzzers, sanitizer and Valgrind-family tools, inverse tests, 16 GiB of stored streams, PractRand, and Dieharder all completed without the anomalies those tests were designed to detect. Full-hash avalanche closely matched the first two moments of an ideal 512-bit binomial model, and state diffusion reached approximately half the 2048-bit state by two rounds.

The conclusion must remain narrow. This is strong implementation evidence for a research prototype, not a security proof. Independent cryptanalysis, reduced-round attacks, formal trail searches, and explicit security-margin arguments are required before any production claim.

\appendix

\section{Parity-Dependent Lane Permutations}\label{app:permutations}
For each voice, output lane $j$ receives the input lane at index $\pi_{r,v}[j]$.
\begin{table}[h]
\centering
\caption{Exact lane permutations.}
\small
\begin{tabular}{ccl}
\toprule
Parity & Voice & $\pi_{r,v}$ \\
\midrule
Even & 0 & 0,3,6,1,4,7,2,5 \\
Even & 1 & 5,0,3,6,1,4,7,2 \\
Even & 2 & 2,5,0,3,6,1,4,7 \\
Even & 3 & 7,2,5,0,3,6,1,4 \\
Odd & 0 & 1,6,3,0,5,2,7,4 \\
Odd & 1 & 4,1,6,3,0,5,2,7 \\
Odd & 2 & 7,4,1,6,3,0,5,2 \\
Odd & 3 & 2,7,4,1,6,3,0,5 \\
\bottomrule
\end{tabular}
\end{table}

\section{Selected Conformance Vectors}\label{app:vectors}
\begin{table}[h]
\centering
\caption{Sequential v0.3 vectors. Hex digests are split across lines for readability.}
\small
\begin{tabularx}{\textwidth}{p{0.14\textwidth}X}
\toprule
Input & Digest \\
\midrule
Empty & \ttfamily 5a6e1f5691b551d2415093151447d619f719ad920bb8061f0f88a586e7a9cf60\\
& \ttfamily b2086bc65d577dbabccedd0a47c0adaeb2523f954ac97afbdb1a01acaa3dd961 \\
\addlinespace
\code{abc} & \ttfamily d73421ee1419c7ba0a5da91d9e728c5b4383e26337dec3c377ca53f656b556b1\\
& \ttfamily 1bc924b8677163e88187a31481f553e2c1c4abc48fe213ce11a1208d71474113 \\
\bottomrule
\end{tabularx}
\end{table}

\begin{table}[h]
\centering
\caption{Tree-mode v0.3 vectors.}
\small
\begin{tabularx}{\textwidth}{p{0.14\textwidth}X}
\toprule
File bytes & Tree digest \\
\midrule
Empty & \ttfamily 8ea132bfaa2f917c7cf0d90eb7912ce7035ad2f568bc9ac8f9af87cc63d86d05\\
& \ttfamily e6cb40d8f8469b47606249a8b789ede703f8217933e299ffe838d31d9b3f4be9 \\
\addlinespace
\code{abc} & \ttfamily 0886b02955bf455af0d7371a8e1bb3bb49afa13a6a8c9241844062da178c55bf\\
& \ttfamily 28942fc0eb1511a5b0e6e62acaea904be0534e9a905a5b277c8874d22d82e5b2 \\
\bottomrule
\end{tabularx}
\end{table}

\section{Complete Reduced-Round Summary}
\begin{longtable}{rrrrrr}
\caption{Insane-profile reduced-round measurements.}\label{tab:all-rounds}\\
\toprule
Rounds & Avalanche mean & Min & Max & Rotational distance & Sampled linear bias \\
\midrule
\endfirsthead
\toprule
Rounds & Avalanche mean & Min & Max & Rotational distance & Sampled linear bias \\
\midrule
\endhead
1 & 588.003 & 219 & 868 & 1024.083 & 0.007385 \\
2 & 1023.946 & 929 & 1115 & 1024.199 & 0.002747 \\
3 & 1023.851 & 946 & 1120 & 1024.273 & 0.002930 \\
4 & 1023.969 & 943 & 1103 & 1023.623 & 0.005432 \\
5 & 1024.128 & 937 & 1114 & 1024.274 & 0.006653 \\
6 & 1023.831 & 938 & 1126 & 1023.898 & 0.003418 \\
7 & 1023.909 & 935 & 1101 & 1024.132 & 0.001648 \\
8 & 1023.821 & 939 & 1111 & 1024.229 & 0.004395 \\
9 & 1023.851 & 932 & 1112 & 1024.106 & 0.003174 \\
10 & 1024.059 & 936 & 1112 & 1023.929 & 0.000122 \\
11 & 1024.086 & 942 & 1107 & 1023.878 & 0.000488 \\
12 & 1024.169 & 931 & 1115 & 1023.916 & 0.001343 \\
13 & 1023.948 & 944 & 1114 & 1024.118 & 0.003784 \\
14 & 1024.198 & 914 & 1120 & 1024.088 & 0.001404 \\
15 & 1023.956 & 940 & 1109 & 1023.977 & 0.003479 \\
16 & 1023.845 & 931 & 1113 & 1024.075 & 0.000244 \\
\bottomrule
\end{longtable}

\section{Compact Pseudocode}
\begin{lstlisting}[caption={High-level sequential hashing skeleton.}]
function FrogBard512(message):
    V = IV
    V = P(V)

    for each complete 128-byte block M:
        V[0][0..7] XOR= LE64(M[0..63])
        V[1][0..7] XOR= LE64(M[64..127])
        V = P(V)

    B = zero[128]
    copy remaining bytes into B
    B[remaining_length] XOR= 0x0b
    B[127] XOR= 0x80
    absorb B and apply P

    V[2][6] XOR= total_length_low64
    V[2][7] XOR= total_length_high64
    V[3][6] XOR= 0x496e2046726f6720
    V[3][7] XOR= 0x7765205472757374

    V[3][0] XOR= 0x21
    V = P(V)
    V[3][1] XOR= 0x2100
    V = P(V)

    return little_endian(V[0][0..7])
\end{lstlisting}


\begin{thebibliography}{99}

\bibitem{bertoni-sponge}
G.~Bertoni, J.~Daemen, M.~Peeters, and G.~Van Assche.
\newblock Sponge functions.
\newblock In \emph{ECRYPT Hash Workshop}, 2007.

\bibitem{keccak-reference}
G.~Bertoni, J.~Daemen, M.~Peeters, and G.~Van Assche.
\newblock The Keccak reference.
\newblock Submission to NIST (Round 3), 2011.

\bibitem{nist-fips202}
National Institute of Standards and Technology.
\newblock SHA-3 Standard: Permutation-Based Hash and Extendable-Output Functions.
\newblock FIPS PUB 202, 2015.

\bibitem{nist-fips180}
National Institute of Standards and Technology.
\newblock Secure Hash Standard (SHS).
\newblock FIPS PUB 180-4, 2015.

\bibitem{aumasson-blake2}
J.-P. Aumasson, S.~Neves, Z.~Wilcox-O'Hearn, and C.~Winnerlein.
\newblock BLAKE2: simpler, smaller, fast as MD5.
\newblock In \emph{Applied Cryptography and Network Security}, LNCS 7954, pp. 119--135, 2013.

\bibitem{blake3}
J.~O'Connor, J.-P. Aumasson, S.~Neves, and Z.~Wilcox-O'Hearn.
\newblock BLAKE3: one function, fast everywhere.
\newblock Technical specification, 2020.

\bibitem{daemen-rijmen}
J.~Daemen and V.~Rijmen.
\newblock \emph{The Design of Rijndael: AES -- The Advanced Encryption Standard}.
\newblock Springer, 2002.

\bibitem{nyberg}
K.~Nyberg.
\newblock Differentially uniform mappings for cryptography.
\newblock In \emph{Advances in Cryptology -- EUROCRYPT '93}, LNCS 765, pp. 55--64, 1994.

\bibitem{bernstein-chacha}
D.~J. Bernstein.
\newblock ChaCha, a variant of Salsa20.
\newblock In \emph{Workshop Record of SASC 2008}, 2008.

\bibitem{boyar-peralta}
J.~Boyar and R.~Peralta.
\newblock A new combinational logic minimization technique with applications to cryptology.
\newblock In \emph{Experimental Algorithms}, LNCS 6049, pp. 178--189, 2010.

\bibitem{merkle-tree}
R.~C. Merkle.
\newblock One way hash functions and DES.
\newblock In \emph{Advances in Cryptology -- CRYPTO '89}, LNCS 435, pp. 428--446, 1990.

\bibitem{nist-sp80022}
A.~Rukhin et al.
\newblock A Statistical Test Suite for Random and Pseudorandom Number Generators for Cryptographic Applications.
\newblock NIST Special Publication 800-22 Revision 1a, 2010.

\bibitem{testu01}
P.~L'Ecuyer and R.~Simard.
\newblock TestU01: a C library for empirical testing of random number generators.
\newblock \emph{ACM Transactions on Mathematical Software}, 33(4), 2007.

\bibitem{valgrind}
N.~Nethercote and J.~Seward.
\newblock Valgrind: a framework for heavyweight dynamic binary instrumentation.
\newblock In \emph{Proceedings of PLDI 2007}, pp. 89--100, 2007.

\bibitem{practrand}
C.~Doty-Humphrey.
\newblock PractRand: practical random number testing, version 0.96.
\newblock Software distribution and documentation.

\bibitem{libfuzzer}
LLVM Project.
\newblock libFuzzer -- a library for coverage-guided fuzz testing.
\newblock LLVM documentation.

\bibitem{aflpp}
AFL++ Project.
\newblock AFL++ documentation and software distribution.

\end{thebibliography}
\end{document}